\documentclass[12pt]{article}
\usepackage{graphicx}
\usepackage{amsmath}
\usepackage{amsfonts}
\usepackage{amssymb}
\begin{document}

\title{Cosmic particles with energies above 10$^{19}$~eV:
a brief review of results\protect\footnote{Talk given at the
session of the Division of Physical Sciences of the Russian Academy of
Sciences dedicated to the 100th anniversary of the discovery of cosmic
rays; to be published in Physics Uspekhi.}}

\author{S.V.~Troitsky\\
\small\it
Institute for Nuclear Research of the Russian Academy of Sciences,\\
\small\it
60th October Anniversary Prospect 7A, \\
\small\it
117312, Moscow, Russia
}
\date{}
\maketitle

\begin{abstract}
Experimental results on ultra-high-energy cosmic rays are briefly
reviewed and their interpretation is discussed. The results related to
principal observables (arrival directions, energies and composition) of
primary particles of extended air showers  as well as
particle-physics implications are addressed.
\end{abstract}

\section{Introduction}
\label{sec:intro}
Ultra-high-energy (above $10^{19}$~eV) cosmic rays (UHECRs) continue to
attract interest of researchers working in both particle physics and
astrophysics for decades. Questions arisen in this field have been related
to the origin of particles with these high energies which do not appear in
the Universe under any other conditions as well as to searches of new
physics which may reveal itself in this energy band and result in
deviation of experimental results from theoretical expectations. As we
will see below, these two groups of questions remain topical and, to a
large extent, determine the present development of research at the border
of particle physics and astrophysics.

Studies of UHECR physics are restricted by two principal complications
related to specific properties of the phenomena under investigation.
Firstly, the flux of these particles is very low (on average, only one
particle with energy we consider arrives at one square kilometer per
year). This implies impossibility of direct registration of primary
particles, which interact in the upper layers of the atmosphere, with the
help of flying detectors, and determines the necessarily indirect way of
their studies with ground-based installations which detect extended
atmospheric showers (EAS) caused by these particles. Moreover, even large
ground-based detectors working for many years collect the number of events
which is negligible as compared, for instance, to the number of
astrophysical photons detected by a telescope in any other energy band.
Secondly, the interaction of the particles with the atmosphere occurs at
energies far beyond the laboratory reach (for a
$10^{19}$~eV proton interacting with an atmospheric nucleon at rest, the
center-of-mass energy is hundreds TeV), therefore the models which relate
the EAS development to properties of the primary particle inevitably
include extrapolation of interaction properties into yet unexplored domains
of energy (and momentum transfer).

\begin{figure}
\begin{center}
\includegraphics[width=0.7\textwidth]{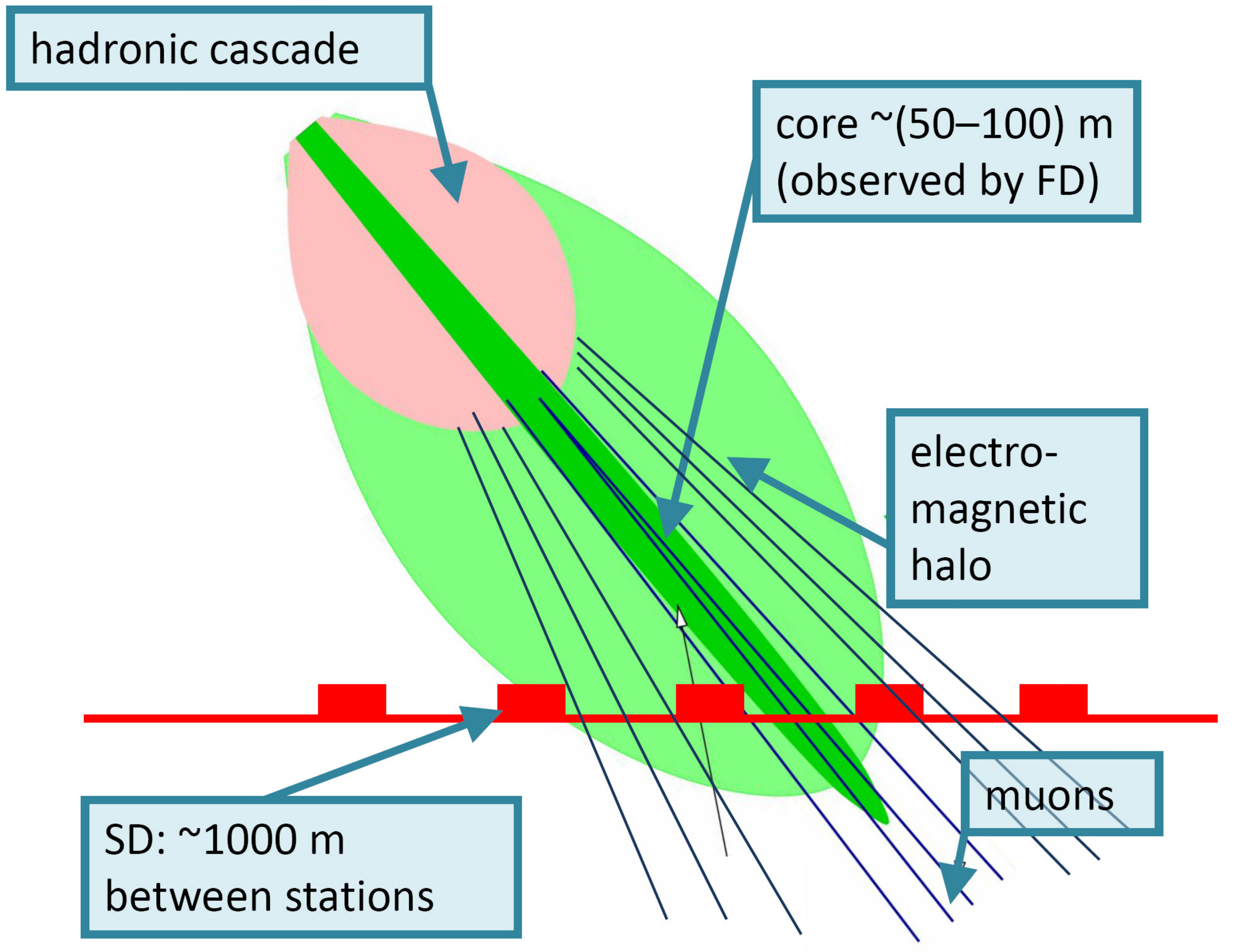}\\
\end{center}
\caption{\label{fig:kolbasa}
\small
A sketch of the EAS development and detection.
}
\end{figure}

The experimental installations actively working at present may be
separated, based on the techniques they use, into ground
arrays of surface detectors (SD) and fluorescent telescope detectors (FD).
SD detects particles of a EAS at the surface level. Detectors form an
array with the spacing $\sim 1$~km and are capable to determine the
lateral distribution function (LDF) of the particle density in the shower.
FD is a telescope which detects ultraviolet emission caused by
fluorescence of atmospheric nitrogen molecules which are excited by
charged particles of the shower. SD registers a two-dimensional slice of
an EAS only but it works independently of the weather conditions and time
of the day and is able to detect various shower components
(electromagnetic, muon, baryon). FD sees the longuitudinal development of
a shower but is able to register events in clear moonless nights only
(roughly, this constitutes about 10\% of time) and is sensitive to the
electron component only. At the same time, SD detects mostly the periferic
part of the shower while FD sees the central core (see
Fig.~\ref{fig:kolbasa}).

Presently, three experiments in the world are capable of studying EAS
caused by primary particles with energies in excess of $10^{19}$~eV. They
are very diferent from each other and have different advantages and
disadvantages.

\textit{ The Yakutsk complex EAS array}
works already for more than 40 years and, presently, have SD of plastic
scintillators covering about 10~km$^{2}$, moderate by the modern
standards. Its principal advantage is the possibility of simultaneous
detection of various EAS components. It is the only modern installation
which provides for large-exposure data of muon detectors; these results
are extremely useful in both the analysis of primary chemical composition
and in testing models of high-energy particle interactions.

\textit{The Telescope Array} (TA) experiment,
located in USA (the state of Utah), is operated by an international
collaboration and
combines an SD of plastic scintillators with the area of array of $\sim
680$~km$^2$ and three FD stations. An important advantage of this
installation is the possibility of hybrid regime, that is of simultaneous
detection of one and the same EAS by both SD and FD with independent
reconstructions (for FD, it can be done in stereo).

The largest modern UHECR experiment, the international \textit{Pierre
Auger Observatory} (PAO) in Argentia, has the area of the SD array of
$\sim 3000$~km$^{2}$, the fact which makes it an undoubted leader in the
exposure, and four FD stations. The observatory also is capable of
hybrid detection, however, the reconstruction by FD is always dependent
from SD and in particular the stereo data are not available. One might
doubt whether the choice of water tanks for SD
was perfect: these detector stations are hypersensitive to the muon EAS
component, the one which is the worst understood in EAS models; this
results sometimes in increased systematic uncertainties.

Together with past experiments which had already finished their work,
these ones obtain sometimes results which are not in a full
mutual agreement. Notably, in 2012, working groups have been created
which include representatives of all three currently operating
experiments. The first results of the work of these groups have been
discussed at a conference in CERN last spring; the discussion in
Sec.~\ref{sec:results} will be based on them in a number of points.

\section{Principal observables}
\label{sec:observables}
In this section, the principal UHECR observables are defined,
those related both to an individual EAS and to the ensemble of data. This
information will be used in the following section where experimental
results are discussed. Independently of the EAS detection method, the
processing of raw data allows to extract the information on a few basic
parameters of the primary particle, namely its type, energy and arrival
direction.

\textbf{Arrival direction.}
The least model dependent observable reconstructed from an EAS is the
arrival direction of the primary particle whose determination is
purely geometrical.  SD reconstructs the arrival direction from the
trigger time of individual detector stations to which the shower
front, moving almost at the speed of light, arrives non-simultaneously. FD
is able to directly fix the position of the plane containing the shower
core and the detector position; in stereo, the core position is given by
the intersection of two such planes; in case of observations by only one
telescope, it is necessary to take into account the temporal development of
the signal. The precision of the SD geometrical reconstruction depends on
the number of triggered stations in addition to the precision of time
measurements; in the FD case the key parameter is the distance between the
telescope and the shower core. In practice, the precision, with which the
arrival direction is determined, decreases with the growth of the
effective area of the detector: SD stations are positioned at larger
spacing and FD telescopes observe a larger volume in the atmosphere. The
best-ever angular resolution (68\% of events reconstructed with the
precision not worse than 0.6$^{\circ}$) has been achived in the
previous-generation experiment, HiRes, which operated two FD stations in
stereo. For present-day experiments with large effective area this
quantity is $\sim 1.5^\circ$.

\textbf{Energy.}
The primary energy is reconstructed indirectly. In the SD case, the signal
is recorded at each particular detector station, then the lateral
distribution of the signal is compared to the expected one. This procedure
of the energy determination introduces considerable uncertainty related to
the modelling of the expected signal for various energies. FD observes the
shower core which carries the dominant part of the energy; this method
allows one to estimate, on the basis of measurements, the total energy of
electrons and positrons in the core and is thus often called
calorimetric. One should note however that there remain significant
sources of uncertainty related both to the value of the fluorescent
yield and to the estimate of the energy not carried by the core electrons.
In all cases, an additional source of (statistical) uncertainty is related
to fluctuations in the first interactions of particles in the atmosphere.
Presently, the energy of a particular primary particle is estimated with
the statistical error of
$\sim (15-20)\%$ and with the systematic uncertainty of $\sim 25 \%$.

\textbf{The type of the primary particle.}
Due to both considerable fluctuations in the development of EAS initiated
by similar primaries and similarities in showers initiated by different
primaries, it is presently hardly possible to determine the type of the
original particle for a particular event. Approaches to this problem are
based on the study of particular EAS components (electromagnetic, muon,
hadron, Cerenkov etc.) and of detailed properties of longuitudinal and/or
lateral shower development (depth of the maximal development, front shape
etc.). Even probabilistic estimates which result from application of these
methods are strongly model-dependent.

\textbf{Observables of an ensemble of EAS.}
Three principal observables determined for each event make it possible to
analyze the ensemble of showers and to obtain statistical information
about UHECR properties, that is about the primary composition, the energy
spectrum and the distribution of arrival directions. For the latter, one
searches for deviations from an isotropic distribution at either large
(global anisotropy) or small (clustering; correlation with potential
sources) angular scales. Results of these studies will be discussed in the
next section.

\section{Review of experimental results}
\label{sec:results}

\subsection{Energy estimation and spectrum}
\label{sec:res:spectrum}
\begin{figure}
\begin{center}
\includegraphics[width=0.7\textwidth]{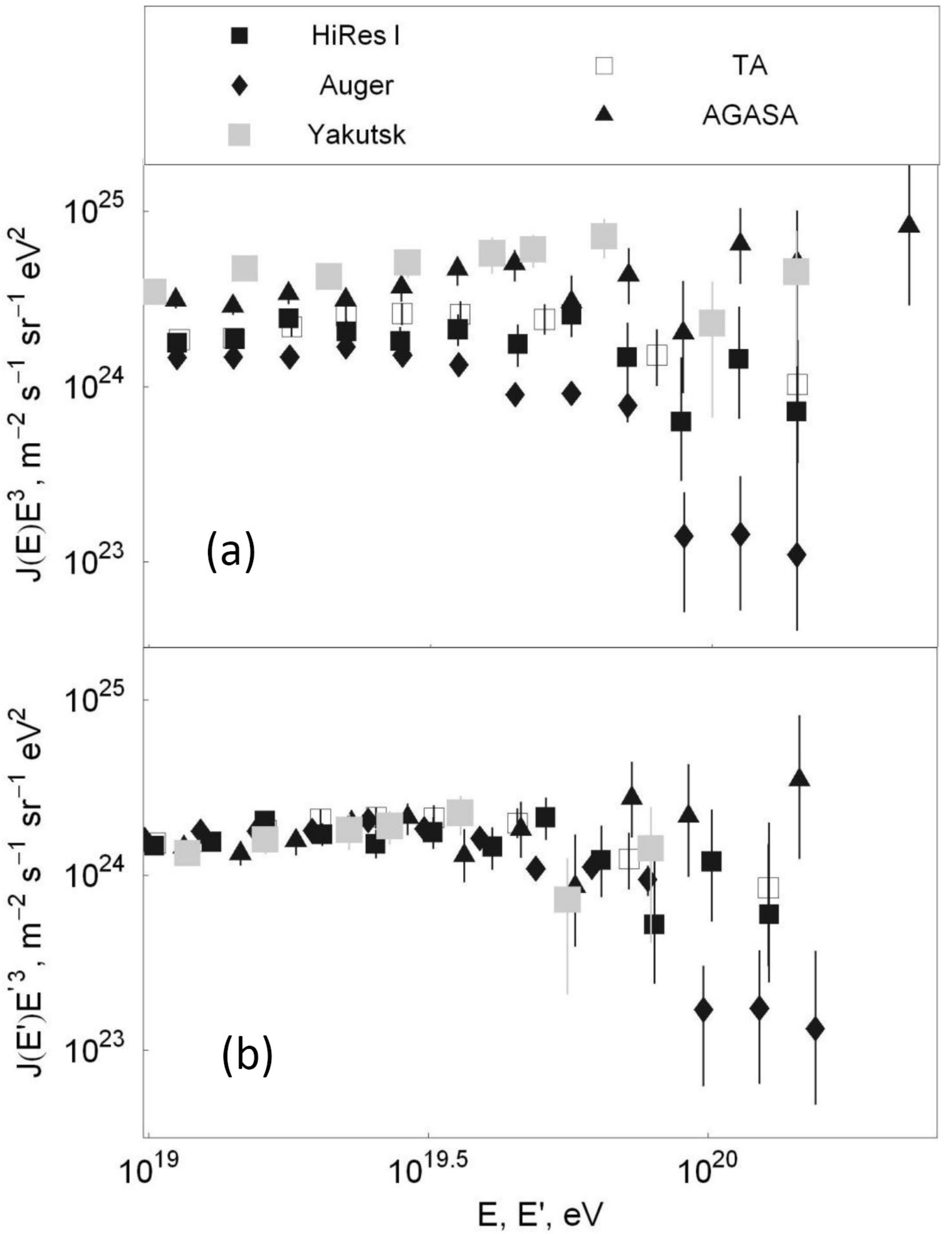}\\
\end{center}
\caption{\label{fig:spectra}
\small
The UHECR spectra (the particle flux $J(E)$) as measured by the
AGASA \cite{AGASA:spectrum},
Yakutsk \cite{Yak:spectrum}, HiRes~I \cite{HiRes:spectrum}, PAO
\cite{PAO:spectrum} and TA \cite{TA:spectrum} experiments, (a)~before and
(b)~after the energy-scale shifts.
The amounts \cite{WG:spectrum} of the energy shifts are
$E'/E=0.652,\,0.561,\,0.911,\,1.102,\,0.906$, correspondingly.}
\end{figure}

UHECR energy spectra measured by various experiments are given in
Fig.~\ref{fig:spectra}(a). Determination of the spectrum which is based on
the absolute measurements of the primary-particle energy and, for FD, also
on detailed simulation of the exposure, cannot be model-independent.
In order to suppress both the arbitraryness related to the choice of model
and the systematic errors, it has been suggested~\cite{Berezinsky} that the
reason for difference of the spectra reconstructed by various experiments
is the energy-independent systematic error of the energy measurement.
Indirectly, this suggestion is supported by the systematic difference
between FD and SD energies for primary particles
of EAS reconstructed by the two methods simultaneously, both in PAO and in
TA. The amount of the related systematic shifts is easy to find by
requiring that the spectra measured by different experiments coincide. To
determine the absolute normalization, one needs an additional theoretical
assumption; in Ref.~\cite{Berezinsky} the energy scale is calibrated by
the theoretically predicted position of a spectral dip related to the
proton energy losses by production of electron-positron pairs.
In a wide energy interval,
$10^{17.5}$~eV$\lesssim E \lesssim
10^{19.5}$~eV,
both the shape and the normalization of the shifted spectra coincide; this
fact strongly supports the approach.
One may however see from Fig.~\ref{fig:spectra}(b) that
this agreement is slightly worse at the highest energies.

For a long time, the interest in the UHECR physics was heated by the
predictions by Greizen~\cite{G}, Zatsepin and Kuzmin~\cite{ZK} of the
cutoff expected in the spectrum of cosmic-ray protons at energies above
$\sim 7\times 10^{19}$~eV which correspond to the pion-production
threshold in proton interactions with photons of the cosmic
microwave background (CMB) radiation (the GZK effect); at the same time,
EAS initiated by particles whose reconstructed energies exceeded
$10^{20}$~eV have been observed experimentally (the first of these events
had been detected by the Volcano Ranch
experiment~\cite{VolcanoRanch:event} even before the CMB was discovered).
As one may see from Fig.~\ref{fig:spectra}, the existence of these events
have been confirmed by all experiments; however, the latest data indicate
the presence of the spectral suppression \cite{HiRes:spectrum, TA:spectrum,
PAO-GZK}. The statistical significance of the suppression is usually
estimated by a comparison of data with the continuation of a power-like
spectrum which is excluded at a certain confidence level. Clearly, the
quantitative estimates of significance depend on the model of the spectrum
continuation; because of that, we do not quote the numbers here. One
should remember also that these results do not prove that the suppression
is related to the GZK effect, nor they exclude a step-like continuation of
the spectrum.

\subsection{Primary composition}
\label{sec:res:composition}
Presently, the question about the UHECR primary composition is open. For a
few recent years, contradictory results of HiRes and PAO are under active
discussion both at conferences and in the literature. While results of the
former experiment are in a full agreement with the
energy-independent mostly proton composition, the measurements by the
latter indicate a graduate change towards heavier primary nuclei with
the energy growth.  Both analyzes used, as the principal observable, the
depth  $X_{\rm max}$  of the maximal shower development, as determined by
FD, and the amount of its fluctuations. Besides these two experiments,
$X_{\rm max}$ has been studied, with smaller statistics, with the FD data
at TA and with the Cerenkov-light data in Yakutsk (in the latter case, the
fluctuations have been also estimated).

The results of all experiments located in the Northern hemisphere (and
therefore observing the Northern sky) agree with the proton composition,
contrary to the PAO (Southern hemisphere) results. This disagreement might
be explained by the presence of nearby sources resulting in a significant
dependence of the primary composition from the direction on the celestial
sphere. However, in 2012, the PAO collaboration presented (see
Ref.~\cite{WG:composition}) a separate analyzis of events arrived from the
Southern and Northern celestial hemispheres (the equatorial part of the
latter may be observed by all experiments); no signs of a systematic
difference were found. The Northern experiments presently have not yet
collected the amount of events sufficient for this kind of analyzis.

Another reason suggested for a possible explanation of the contradiction
in $X_{\rm max}$ results is the difference in methodics of the data
processing by PAO and Yakutsk versus HiRes and TA. While the value of
$X_{\rm max}$ of an individual shower is defined in a similar way by all
groups, the study of the ensemble of showers proceeds differently: the
former pair of experiments select, by means of imposing numerous cuts
which reduce the number of events significantly, the most representative,
minimum-bias sample in which the $X_{\rm max}$ distribution should
coincide with that of all EAS, both detected and missed in the sample.
Contrary, the second pair consider the full set of detected EAS but take
the selection effects into account in calculation of the theoretically
expected values, $X_{\rm max}'$, for a given particular sample. To add to
the complication, HiRes made use of a slightly different, compared to PAO
and Yakutsk, quantity which parametrizes fluctuations. The direct
comparison of results obtained by various experiments is therefore
possible only in terms of the final result, the primary nuclear
composition, which is traditionally parametrized by the mean logarithm of
their athomic mass, $\langle \ln A \rangle$. Unfortunately this analyzis
inevitably depends on the shower-development model which is used to relate
observable parameters to $\langle \ln A \rangle$.

\begin{figure}
\begin{center}
\includegraphics[width=0.7\textwidth]{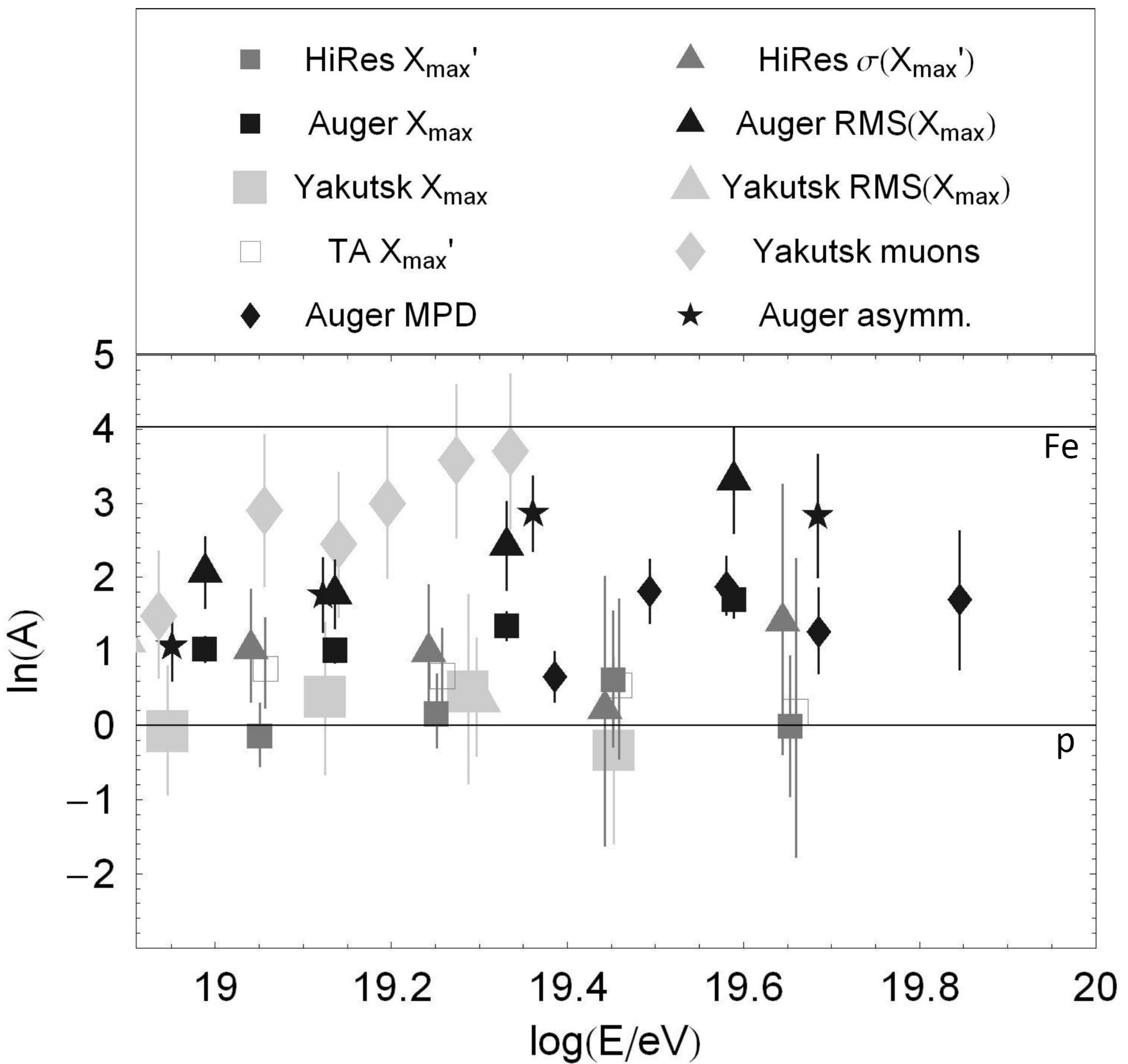}\\
\end{center}
\caption{\label{fig:composition}
\small
Results of the analyses of UHECR primary composition by various methods:
maximal shower development depth and its fluctuations from the data of the
HiRes \cite{HiRes-Xmax},
PAO \cite{PAO-Xmax}, TA \cite{TA-Xmax} and Yakutsk \cite{Yak-Xmax}
experiments;
Yakutsk muon data \cite{Yak-mu}; reconstructed muon production depth
(``MPD'') and shower front shape asymmetry
(``asymm.'') from the PAO SD data \cite{PAO-comp-SD}.}
\end{figure}

The results of this comparative analyzis, which made use of the EAS
parameters mentioned above as well as of some others, are presented in
Fig.~\ref{fig:composition}, where we used QGSJET~II  \cite{QGSJETII}  as a
model of high-energy hadronic interactions (this choice was determined by
the availability of published data for comparison with this model).
In my opinion, the scatter of values of $\langle
\ln A \rangle$ obtained by means of various methods indicates
that it may be too early to claim any significant contradiction
between experiments. In particular, once expressed in terms of $\langle
\ln A \rangle$, the difference in $X_{\rm max}$ results between the HiRes
and PAO analyses does not exceed the difference between PAO $X_{\rm max}$
and fluctuation results. Probably one should state that systematic
errors still dominate over real effects in the studies of the primary
composition.

To finalize the discussion of the chemical composition, note that,
astrophysically, a significant amount of primary heavy nuclei looks less
probable as compared to the (predominantly) proton composition since it
requires additional mechanisms of increasing metallicity in the injected
matter by several
orders of magnitude with respect to the maximal known stellar metallicity.
The argument that particles with larger electric charge are accelerated
more efficiently leads to the requirement of a (not observed
experimentally) sharp jump both in the composition and in the full flux of
cosmic particles at the energies which correspond to the maximal energy of
accelerated protons.

\subsection{(An)isotropy of the arrival directions}
\label{sec:res:anisotropy}
Small number of events, relatively bad angular resolution and deflections
of charged particles by cosmic magnetic fields make it impossible,
presently, to identify UHECR sources object-by-object as it is customary in
the classical astronomy. Instead, one has to operate by statistical
methods and to search for manifestations of particular models of the
population of sources in anisotropic distribution of the cosmic-ray
arrival directions for the entire sample. One may distinguish the searches
for global and small-scale anisotropy.

\textit{The global anisotropy of arrival directions} is expected in the
case when the observed cosmic-ray flux is due to a limited number of
more or less nearby sources. This picture is relevant for two cases:
either there is a significant overdensity of sources close to the observer
or particles from distant sources do not reach us for some reason. The
first case corresponds to the sources in our Galaxy. The second option
takes place for astrophysical sources of protons with sub-GZK energies;
the dominant contribution to the cosmic-ray flux at these energies should
come from the sources inside the so-called GZK sphere with the radius of
order 100~Mpc. Since the matter inside this sphere is distributed
inhomogeneously, the astrophysical scenario with a large number of proton
sources implies anisotropic distribution of the arrival directions. This
distribution may be predicted from a model of the distribution of
sources, that is of matter in the Universe, supplemented by some
assumptions about particle propagation.  On the other hand, searches for
manifestations of some particular classes of sources in
\textit{small-scale anisotropy} consist of, basically, studies of the
autocorrelation function (clustering) or of correlations of cosmic-ray
arrival directions with positions of objects of a certain class.

\begin{figure}
\begin{center}
\includegraphics[width=0.8\textwidth]{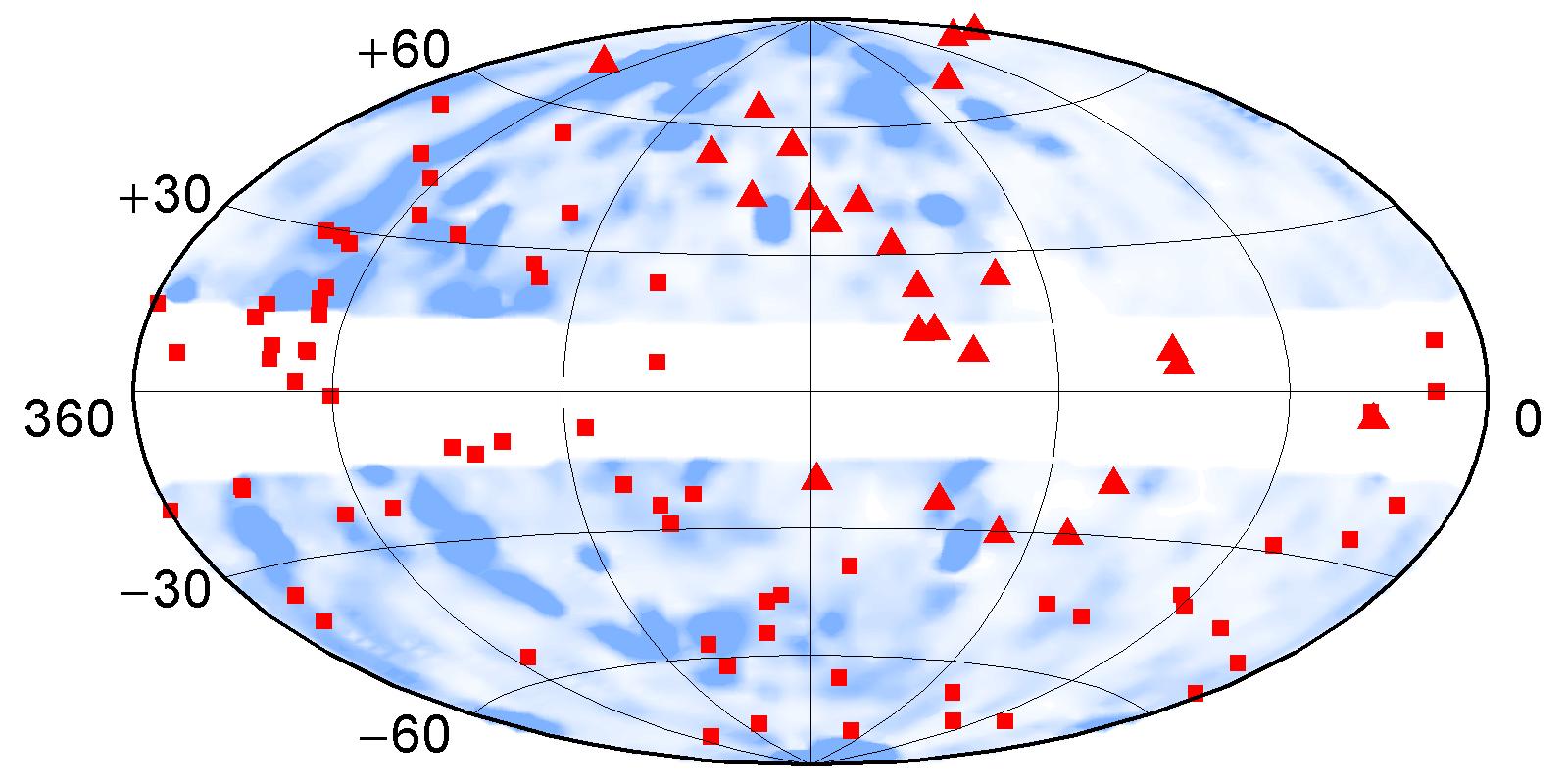}\\
\end{center}
\caption{\label{fig:LSS}
\small
The expected flux of protons with energies $E \gtrsim 5.5\times
10^{19}$~eV from extragalactic sources whose distribution follows the
large-scale structure of the Universe, with the exposures of PAO and TA
taken into account (Galactic coordinates; darker regions correspond to
higher flux; the method of calculation is described in
Ref.~\cite{TUS-aniso}; relative exposures normalized to the number of
events; the white strip corresponds to the zone of Galactic absorption
where precise data on the structure are missing), together with the
arrival directions of PAO~\cite{PAO-AGN-new} (squares) and
TA~\cite{TA-aniso} (triangles) events. }
\end{figure}

Results of most analyses of the distribution of arrival directions of
primary particles with energies above $10^{19}$~eV are in statistical
agreement with the isotropic distribution at a good confidence level. At
the same time, in some particular cases, there are indications to
deviations from isotropy, that is the data, being compatible with the
isotropic distribution, does not exclude some anisotropy scenarios. For
instance, in the Southern hemisphere (PAO), the global distribution of the
arrival directions suggests their possible correlation with the
large-scale structure of the Universe while this is not seen in the data of
Northern experiments (see Fig.~\ref{fig:LSS}); TA results exclude this
correlation at the 90\% confidence level for events with energies
$E>10^{19}$~eV (for $E>4\times 10^{19}$~eV, arrival directions  are
consistent with both scenarios).

One of the recent results most important for astrophysics is the lack of
statistically significant clustering of arrival directions at small
scales. The search for clusters of events allows one to constrain the
number of their sources in the nearby Universe: in the limit when there is
only one source, the arrival directions would all concentrate in a single
spot around it; contrary, for infinite number of sources, the
distribution would be isotropic. A quantitative method, which results in a
lower limit on the number of sources from the lack of clustering, has been
developed in Ref.~\cite{DTT}; its somewhat more complicated version has
been recently applied to the PAO results \cite{PAO-clustering}. Reliable
constraints on the number density of sources may be obtained for the
highest energies where the flux is dominated by nearby sources (because of
the GZK effect) while particle deflections by magnetic fields are not
large. The result of this analysis is the bound
$n \gtrsim 10^{-4}$~Mpc$^{-3}$ on the concentration of sources of
particles with $E \gtrsim 5.5 \times 10^{19}$~eV (under the assumption of
small deflections). It is a very restrictive bound: the sources should be
much more abundant than it is assumed in most theoretical models. Indeed,
simple bounds on the physical parameters of a source of particles with
these energies~\cite{SourcesI} demonstrate that for classical mechanisms
of diffusive acceleration (e.g.\ in shock
waves), the required conditions are fulfilled only in very exotic and rare
oblects, the most powerful active galaxies. At the same time, a less
popular mechanism of direct acceleration of particles in magnetospheres of
supermassive black holes~\cite{Neronov2} allows one to satisfy the
concentration bounds and to construct a model of the population of
sources~\cite{population}.

The autocorrelation function for the arrival directions of events with
$E>10^{19}$~eV is fully consistent with that expected for isotropic
distribution \cite{TA-aniso}; however, at higher energies, slight
deviations from isotropy are observed which consist of excesses of events
separated by the angular scale about 15$^{\circ}$. In the PAO data, this
excess is determined by a spot of events \cite{Comment, Comment2} around a
nearby radio galaxy Cen~A. It may happen that this spot is responsible
also for the effect of the correlation with the large-scale structure
because Cen~A is projected to a more distant but very numerous
supercluster of galaxies. In the Northern hemisphere (TA), one does not
see an evident spot but the excess in the autocorrelation function is
present. Note that for $E>10^{20}$~eV, the PAO and TA experiments have
detected 6 events only, of which two coincide within the angular
resolution \cite{doublet}.

One of the best-known results of comparison of particle arrival directions
with positions of astrophysical objects of a certain class is the
conclusion of the Pierre Auger collaboration~\cite{PAOagn} about the
correlation of arrival directions of particles with
$E>5.6 \times 10^{19}$~eV with positions of nearby active galaxies which
has been interpreted as an evidence that the events in this energy range
are caused by protons either from these galaxies or from other objects
distributed in the Universe in a similar way. This conjecture has a number
of problems and is hardly consistent with the analyses of other
observables (including the chemical composition and the global anisotropy)
and with astrophysics of the sources. It has been confirmed by the Yakutsk
data~\cite{YakAGN} and not confirmed by HiRes~\cite{HiResAGN}. More recent
PAO data \cite{PAO-AGN-new} point to a much weaker, as compared to
Ref.~\cite{PAOagn}, effect. The TA results~\cite{TA-aniso} exclude the
effect in its original strength \cite{PAOagn} and are consistent both with
the total absence of the effect and with the weak effect seen in the
latest data \cite{PAO-AGN-new}.

\section{Particle-physics applications}
\label{sec:particle-physics}
It were the cosmic rays which made it possible to discover many elementary
particles in the past, and presently fundamental physics of particles
and interactions continues to exploit information coming from
cosmic-ray physics and astrophysics. The primary directions here
are: to study hadronic interactions at energies an order of
magnitude higher than achieved in accelerators; to search for unknown
effects which influence the atmospheric shower development; to search for
``new physics'' in order to solve problems with the standard
explanation of astrophysical results.

\subsection{Particle interactions at very high energies}
\label{sec:part:interactions}
\begin{figure}
\begin{center}
\includegraphics[width=0.95\textwidth]{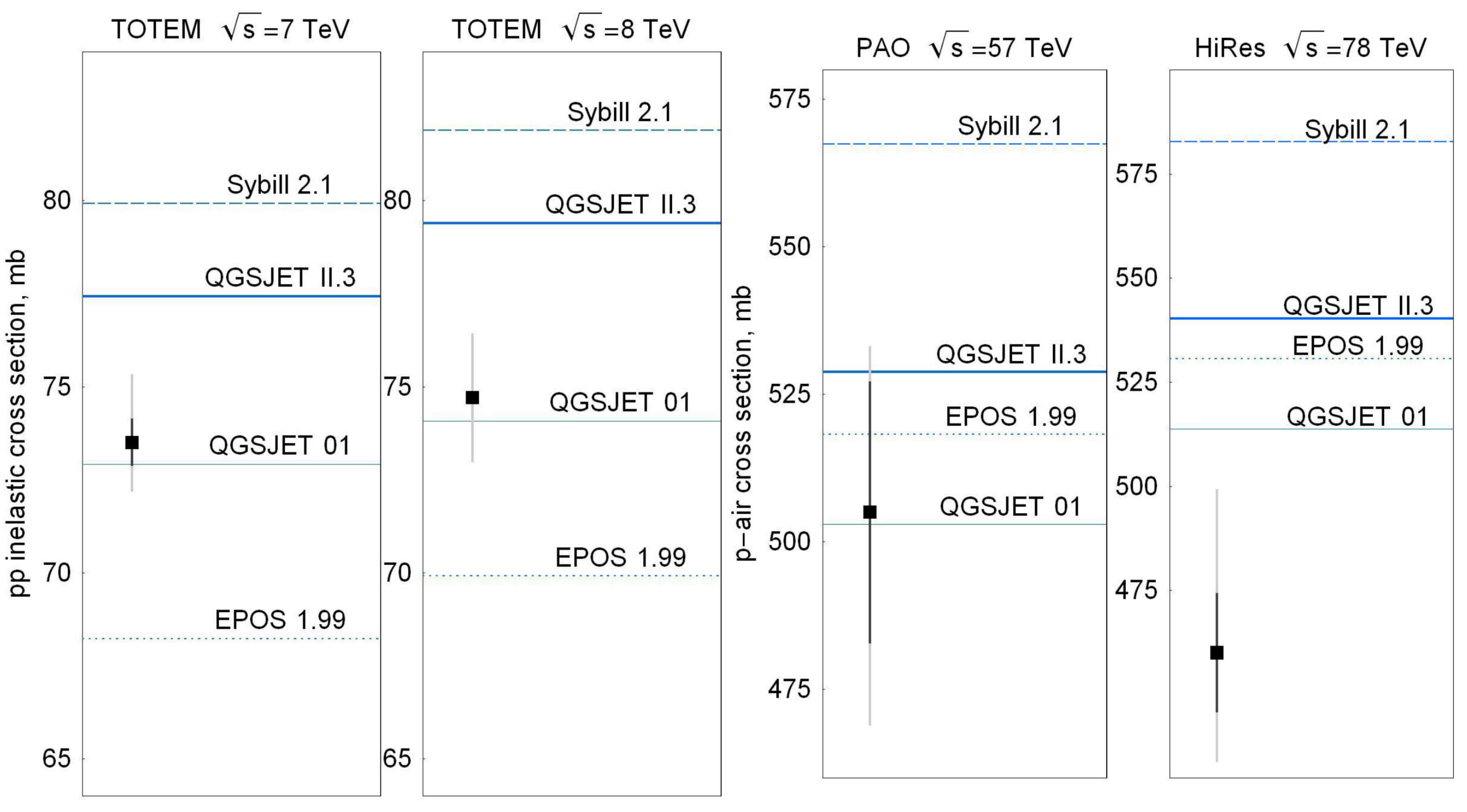}\\
\end{center}
\caption{\label{fig:TOTEM}
\small
A comparison of cross sections used in the hadronic-interaction models
Sybill~2.1 \cite{Sybill},
QGSJET~01 \cite{QGSJET01}, QGSJET~II~\cite{QGSJETII} and EPOS~1.99
\cite{EPOS} (values taken from Ref.~\cite{PAO-cross-sec}),
with the experimental results, left to right:
inelastic $pp$ cross section from the models and from the TOTEM experiment
data at the LHC energy, $\sqrt{s}=7$~TeV \cite{TOTEM} and
$\sqrt{s}=$8~TeV  \cite{TOTEM8}; ``$p$-air'' cross section from the
models and from the EAS analysis data by the PAO \cite{PAO-cross-sec},
$\sqrt{s}=$57~TeV, and HiRes \cite{HiRes-cross-sec},
$\sqrt{s}=$78~TeV, experiments. Statistical and systematic
uncertainties are shown in black and gray, respectively. }
\end{figure}

The center-of-mass energy of a proton-proton collision at the Large Hadron
Collider (LHC) is an order of magnitude less than that of the first
interaction of a UHE particle in the atmosphere. On the one hand, this
results in a large uncertainty in models describing EAS (though LHC
results, in particular those of a dedicated experiment LHCf, are presently
in active use to improve the models, one cannot avoid extrapolation). On
the other hand, measurement of model-independent EAS properties allows to
extract directly quantitative information about the first interaction.
Both aspects are illustrated by Fig.~\ref{fig:TOTEM}.
Today, the precision of both the models and the measurements is
insufficient to make any statement about the influence of new physics on
the shower development.

\subsection{New-physics searches}
\label{sec:part:new-physics}
Let us turn now to a couple of (far from being unique but, in the author's
opinion, most interesting for today) examples of application of UHECR to
the search and constraining of new-physics models, that is of particles
and interactions assumed in theories which extend the Standard Model of
particle physics (SM) and attempt to solve some of its problems
\cite{ST-UFN}.

\textbf{Neutral particles from BL Lac type objects.}
In 2004, the analysis of a data sample with the best ever angular
resolution in UHECR physics (HiRes stereo \cite{HiRes:271events}) revealed
\cite{BL:HiRes} statistically significant correlations of arrival
directions of a small fraction (about 2\%) of cosmic particles with
energies above $10^{19}$~eV with bright BL Lac type objects, powerful
active galaxies of a certain class located far away from the Earth. The
angular resolution of the experiment was much smaller than the value of
the expected deflection of protons with these energies in the Galactic
magnetic field, so this observation pointed to the existence of
UHE \textit{neutral particles} which travel for cosmological distances.
A subsequent publication by the HiRes collaboration~\cite{HiRes:BL}
confirmed this result with an alternative analysis method. This
phenomenon cannot be explained within the frameworks of standard
physics and astrophysics (see e.g.\ the discussion in
Ref.~\cite{TT:neutrals}). Popular extensions of SM, e.g.\ supersymmetry,
do not help as well. The only non-contradictory explanation of this
effect, which helps also to solve some other astrophysical problems and
may be tested experimentally, has been proposed in Ref.~\cite{BL-axion} and
is based on the phenomenon of axion-photon oscillations. Unfortunately, the
very effect has not yet been tested in a similar independent experiment:
because of the worse angular resolution of the only installation (TA)
which operates FD in the stereo mode, one needs a very large number of
events not yet collected. The absence of the effect in the PAO SD data
\cite{Auger:BL} agrees with the predictions of the axion-photon
conjecture: the PAO water tanks are almost insensitive to muon-poor EAS
initiated by primary photons.

\textbf{Superheavy dark matter.}
One of the experimental results which requires SM to be extended for its
explanation is the presence in the Universe  of a large amount of
invisible matter, the so-called dark matter. In a certain class of models
it is supposed that this matter consists of metastable (with lifetime
$\tau _{X}$ of order the lifetime of the Universe) superheavy (mass
$M_X>10^{20}$~eV) particles $X$, among whose decay products UHECR primary
particles may be present. The decay of the $X$ particles may be described
in a sufficiently model-independent way because the key role in its
physics is played by relatively well understood hadronisation processes.
Among the predictions of this scenario are a very hard spectrum at the
highest energies, a large fraction of primary photons and the Galactic
anisotropy of the arrival directions. The most restrictive constraints on
this scenario come presently from the limits on the photon flux but still
leave open~\cite{Kalashev-SHDM} a significant part of the
$X$-particle parameter space. This model attracts some special interest
now because no candidate for the dark-matter particle was found at LHC.

\section{Conclusion}
\label{sec:concl}
The UHECR physics remains, for decades, one of the most interesting fields
at the intersection of astrophysics and particle physics. Despite a
serious progress in the experiment, we presently cannot say a lot about
the origin of particles with energies above
$10^{19}$~eV, and only a few models of particle acceleration in
astrophysical sources may simultaneously satisfy both the constraints on
physical conditions in these accelerators and the strict lower bound on
the number density of sources obtained recently from the absence of
clustering of arrival directions. The results of the studies of chemical
composition of primary particles in this energy range are probably
dominated by systematic errors and not by real physical effects. The
physical reason of a systematic difference in the primary energy
determination by means of different methods is still unknown. Some
indications to possible manifestations of new physics in cosmic rays
deserve a close attention.

This work has been supported in part by RFBR (grants
10-02-01406, 11-02-01528 and 12-02-01203),  the RF President
(grant NS-5590.2012.2), the RF Ministry of Science and Education
(agreements 8142 and 14.B37.21.0457) and the Dinasty Foundation.

\end{document}